\documentclass[3p]{elsarticle}
\usepackage{amssymb,amsmath}
\newcommand{\angstrom}{\textup{\AA}}
\usepackage{graphicx}
\usepackage{cancel}
\usepackage[ruled]{algorithm2e}
\usepackage{siunitx}

\usepackage[colorinlistoftodos,textwidth=4cm,shadow]{todonotes}

\newcounter{Igor}

\date{}
\title {\textbf{Computational Study of Ultrathin CNT Films with the Scalable Mesoscopic Distinct Element Method}}

\author{Igor Ostanin \footnote{Corresponding author, tel: +79150174677, e-mail:i.ostanin@skoltech.ru}
}

\address{Skolkovo Institute of Science and Technology, Nobel St. 3, Moscow, Russia}

\author{Traian Dumitric\u{a}}

\address{University of Minnesota, 111 Church st., Minneapolis, USA}

\author{Sebastian Eibl, Ulrich R{\"{u}}de}

\address{Friedrich-Alexander University Erlangen-Nuremberg, Cauerstr.11, Erlangen, 91052, Germany}

\begin{document}

\begin{abstract}

In this work we present a computational study of the small strain mechanics of freestanding ultrathin CNT films under in-plane loading. The numerical modeling of the mechanics of representatively large specimens with realistic micro- and nanostructure is presented. Our simulations utilize the scalable implementation of the mesoscopic distinct element method of the waLBerla multi-physics framework. Within our modeling approach, CNTs are represented as chains of interacting rigid segments. Neighboring segments in the chain are connected with elastic bonds, resolving tension, bending, shear and torsional deformations. These bonds represent a covalent bonding within CNT surface and utilize Enhanced Vector Model (EVM) formalism. Segments of the neighboring CNTs interact with realistic coarse-grained anisotropic vdW potential, enabling relative slip of CNTs in contact. The advanced simulation technique allowed us to gain useful insights on the behavior of CNT materials. In particular, it was established that the energy dissipation during CNT sliding leads to extended load transfer that conditions material-like mechanical response of the weakly bonded assemblies of CNTs.

\end{abstract}

\maketitle

\section{Introduction}

Ultrathin freestanding carbon nanotube (CNT) films \cite{Wu_Science_2004} possess a number of unique physical and mechanical properties, making them attractive for a number of interesting applications, in particular, in flexible electronics: transparent displays \cite{Park_Nanoscale_2013}, mirrors for soft X-ray radiation \cite{Medvedev_Nanoscale_2019}, wearable medical devices \cite{Wang_SMS_2017} \textit{etc}. The major advantage of these materials, as compared to possible competitors, is their ability to withstand significant plastic deformations without irreversible structural changes. This property is conditioned by discontinuous nature of CNT films - they consist of individual CNTs that are chemically reactive and bound only by van der Waals forces. Due to molecular structure of CNTs, the interfaces between neighboring parallel CNTs resolve relative sliding of one CNT along the other, with very small static friction. Such sliding provides mechanism for plastic flow of CNT material. This mechanism is non-destructive in a sense that small strain plastic deformations do not lead to significant changes in microstructure of a CNT material, as well as in its mechanical, thermal, optical and electric properties. Another interesting feature is that such plastic deformations do not immediately localize in a small damage zone, but for a considerable strain diapason remain distributed over the length of a tested specimen.

The modern capabilities for experimental research on the properties of CNT materials are limited both in terms of technology and cost efficiency. Thus, it is highly desirable to have a numerical model that could reliably reproduce the mechanics of thin CNT films. Although molecular-level modeling techniques are capable to provide useful insights on the properties of individual nanotubes \cite{Yakobson_PRL_1996,Dumitrica_PNAS_2006,Zhang_APL_2008,Nikiforov_APL_2010} they're inefficient for modeling assemblies comprising large numbers of CNTs - the most extreme molecular dynamics calculations so far included only hundreds of CNTs \cite{Cornwell_JCP_2011}.

In order to bypass the computational inefficiency of full molecular modeling techniques, a number of coarse-grained mesoscale models were developed \cite{Buehler_JMR_2006,Cranford_Nanotech_2010,Mirzaeifar_Nanoscale_2015,Volkov_ACS_2010,Volkov_JPC_2010,Volkov_PRL_2010,Wittmaack_CST_2018}. The general idea of these models is to reduce the number of model degrees of freedom (DOF) in order to reach the computationally tractable model size. However, such coarse-graining often leads to significant artifacts in model's behavior. For example, pair vdW potential used in \cite{Buehler_JMR_2006,Cranford_Nanotech_2010} leads to absence of CNT relative sliding, which makes the model inapplicable for modeling irreversible deformations of CNT assemblies. Similarly, giving up on torsional degrees of freedom in models \cite{Volkov_ACS_2010, Volkov_JPC_2010} significantly alters the mechanical properties of CNT assemblies under certain types of loading. Moreover, neither model so far provides a realistic description of energy dissipation during relative sliding of CNTs, which makes it impossible to describe rate-dependent deformation properties of CNT materials.

In our works \cite{Ostanin_JMPS_2012,Ostanin_JAM_2014,Ostanin_JMR_2015} we have developed the mesoscopic modeling technique based on distinct element method \cite{Cundall_Geo_1979}. Within our approach, each CNT is represented as a chain of rigid bodies possessing tensile, bending, shear and torsional stiffnesses. These stiffnesses are calibrated based on molecular dynamics simulations. CNTs interact via vdW potential tailored to provide adequate description of relative CNT sliding. In our athermal model, energy dissipation was represented with viscous damping, appropriately calibrated in a top-down fashion \cite{Ostanin_JAM_2014}. The model showed to be efficient for the description of CNT self-folded configurations \cite{Ostanin_APL_2013,Ostanin_SoftMatter_2014}, weakly bonded and cross-linked CNT ropes and bundles \cite{Ostanin_JAM_2014,Ostanin_JMR_2015} and qualitative description of large-strain deformations of CNT films \cite{Wang_CAR_2018}. Until recently, the model was lacking a scalable parallel implementation, which severely limited its capabilities in terms of achievable length and time scales of the simulation. We addressed this reservation with the development of a scalable parallel implementation of our model based on the rigid particle dynamics module of the waLBerla package \cite{Ostanin_LOM_2018}.    

In our present work, utilizing our new parallel computation capabilities, we investigate the mechanics of representatively large assemblies of thin films. We present the study of the representative volume element (RVE) size, dependence of the mechanics of CNT films on dissipation rate, film thickness and length of individual CNTs. We restrict our consideration with the case of small strain deformation of CNT materials, with strains less than few percent. We could see that even at such strains CNT films exhibit complex nonlinear behavior, conditioned by CNT sliding within the tested specimen. Our observations are in good agreement with recent experimental studies on the mechanics of CNT films.

\section{Method}

Our mesoscopic model mostly corresponds to the ones employed in our earlier works \cite{Ostanin_JMPS_2012,Ostanin_JAM_2014,Ostanin_LOM_2018}. For clarity, we present a brief description of our modeling technique here. The model is based on the Mesoscopic Distinct Element Method (MDEM), that computes the damped dynamics of a collection of interacting rigid bodies of given mass, shape and tensor of inertia -- distinct elements. The state variables for each element include translational positions and velocities, as well as rotational positions and angular velocities. The bodies change their velocities and angular velocities due to contact forces and moments arising in pair interactions, as well as external forces and moments, acting at each body. The system is evolved in time with an explicit velocity Verlet time integration scheme.

\begin{figure}
	\includegraphics[width=16cm]{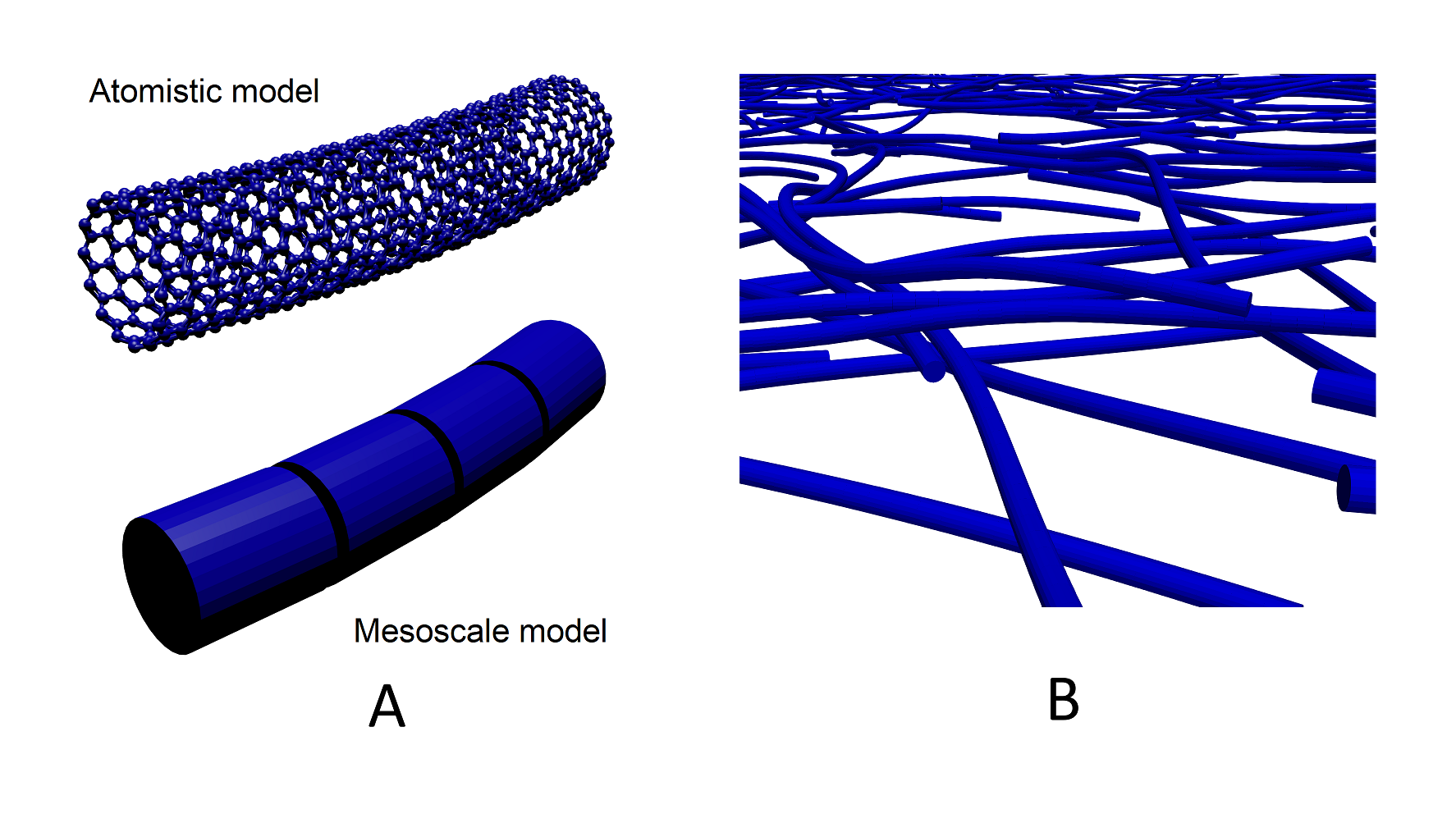}
	\protect\caption{(A) Representation of a CNT within our coarse-grained model (B) Coarse-grained representation of multiple interacting CNTs}
\end{figure}

Within our approach, undeformed CNTs are homogenized into cylindrical shells with finite thickness and then partitioned into identical segments of finite lengths $T$ (Fig. 1). Each distinct element represents the inertial properties of a CNT segment. Table 1 provides segment parameters for $(10,10)$ CNTs with diameter $2r_{CNT}=13.560$ $\mathring{A}$ and length $T=2r_{CNT}$. Each segment contains approximately $220$ carbon atoms.

The inertial properties of a segment are represented with mass $m$ and the moment of inertia $I = \frac{2}{5}mr^2$ (a \textit{spherical} tensor of inertia of a distinct element is assumed here for simplicity, however, a number of non-spherical shapes are immediately available in the framework). Parameters $m$ and $I$ are matched with the mass and moment of inertia of a cylindrical segment taken with respect to the CNT axis by proper choice of the radius of a distinct element:
\begin{equation}
r=\sqrt{2.5}r_{CNT}.  \label{4}
\end{equation}
The elements representing CNT segments are equispaced at a distance $T$.

Such a coarse-graining leads to the elimination of atomistic degrees of freedom. Therefore, the dissipative microscopic
processes associated with CNT sliding are not explicitly
captured and should be included in a phenomenological manner. In this respect, two independent channels of energy dissipation are introduced in our model -- local damping and viscous damping, controlled by the parameters $\alpha$ and $\beta$ respectively.   

PFC-style \textit{local damping} \cite{Itasca_2008} acts at each body. It is introduced to damp stiff interactions and stabilize time integration in dense particle assemblies. The components of damping force $F^{\alpha}_i$ (moment $M^{\alpha}_i$ ) are proportional to the corresponding components of unbalanced force $F_i$ (moment $M_i$) according to:

\begin{equation}
F^{\alpha}_i = - \alpha \left|F_i\right| sign(v_i), M^{\alpha}_i = - \alpha \left|M_i\right| sign(\omega_i)   \label{1}
\end{equation}

Here $v_i$ and $\omega_i$ are components of the translational and rotational velocity of an element, and $sign(x)$ is the sign function. In our simulations the local damping is kept relatively low ($\alpha=0.1$) and its influence on CNT dynamics is insignificant.

Recent theoretical findings \cite{Brilliantov_2017} predict a linear dependence between force and velocity in nanoscale thermal-induced friction. In our coarse-grained model we introduce the viscous damping that provides a simplified model for nanotribological energy losses during CNT sliding.  The viscous damping forces, proportional to relative segment velocities, act in parallel with vdW contact forces. These forces are controlled by parameter $\beta$. Normal $F^{\beta}_n$ and tangential $F^{\beta}_s$ viscous forces are proportional to normal $v_n$ and tangential
$v_s$ relative velocities of elements in vdW contact:

\begin{equation}
F^{\beta}_n = c_n v_n, F^{\beta}_s = c_s v_s,   \label{2}
\end{equation}

Viscosity coefficients $c_n$ and $c_s$ are related to $\beta$ as follows:

\begin{equation}
c_n = 2\beta \sqrt{m k_n}, c_s = 2\beta \sqrt{m k_s},   \label{3}
\end{equation}

where $k_n$ and $k_s$ are linearized stiffnesses of the contact model, taken $k_n = 100 eV/nm^2$, $k_s = 100 eV/nm^2$.

Damping factor $\beta$ is kept as a free parameter in our simulations. Initial estimate of $\beta$ is performed in a top-down way, as discussed in \cite{Ostanin_JAM_2014}. Unless otherwise noted $\beta = 0.1$ is assumed in this work.

Elasticity of CNTs in our model is represented with the formalism of enhanced vector model (EVM) \cite{Kuzkin_PRE_2012}. Unlike incremental parallel bonds, widely used in DEM community \cite{Potyondy_2004}, EVM bonds explicitly conserve energy in symplectic time integration. EVM is based on a binding potential, describing the behavior of an elastic bond linking two rigid bodies. The formulation provides straightforward generalization for the case of large strains and accounts for a bending-twisting coupling. 

Consider two equal-sized particles $i$ and $j$ with equilibrium separation $T$ and orientation described in terms of orthogonal vectors  $n_{ik}$, as depicted in fig. 2(A). Then the EVM bond potential is given as follows:

\begin{equation} \label{5}
U(r_{ij},{n}_{ik},{n}_{jk})=\frac{B_{1}}{2}(r_{ij}-T)^{2}+\frac{B_{2}}{2}(\mathbf{n}_{j1}-\mathbf{n}_{i1})\mathbf{r}_{ij}/{r}_{ij}+B_{3}\mathbf{n}_{i1}\mathbf{n}_{j1}-\frac{B_{4}}{2}(\mathbf{n}_{i2}\mathbf{n}_{j2}-\mathbf{n}_{i3}\mathbf{n}_{j3})
\end{equation}

Here $\mathbf{r}_{ij}$  is the radius vector connecting centers of bonded particles. Parameters  $B_1...B_4$ are directly related to longitudinal, shear, bending, and torsional rigidities of a bond, according to Euler-Bernoulli beam theory (see \cite{Kuzkin_PRE_2012} for more details):

\begin{eqnarray} \label{6}
B_{1}=\frac{ES}{T}, \notag \\
B_{2}=\frac{12EJ}{T},   \notag \\
B_{3}=-\frac{2EJ}{T} - \frac{GJ_{p}}{2T},  \notag  \\
B_{4}=\frac{GJ_{p}}{T}. \notag  \\
\end{eqnarray} 

where $S,I,J$ are area, moment of inertia and polar moment of inertia of a cylinder shell beam with radius $r_{CNT}$ and thickness $h$:

\begin{eqnarray} \label{7}
S=2\pi hr_{CNT}, \notag \\
J=\pi hr_{CNT}(r_{CNT}^{2}+h^{2}/4), \notag \\
J_{p}=2J. \notag \\
\end{eqnarray} 

Table 1 summarizes the elastic properties of CNT segments.

The potential \ref{5} with calibrations \ref{6} reproduces the elastic behavior of a CNT in four admissible deformation modes (fig. 2(B)).

\begin{table}[t] \centering
	\label{t1}  
	\caption{Parameterization of the spherical particles and EVM bonds for a $(10,10)$ CNTs. $m$, $r$, $I$  are the mass, radius, moment of inertia of each spherical particle. $B_1$,$B_2$,$B_3$,$B_4$ are EVM stiffnesses.}
	
	\begin{tabular}{ccccccccc}
		\hline
		$m$ & $r$ & $I$ & $B_1$ & $B_2$ & $B_3$ & $B_4$ \\ \hline
		$(amu)$ & $(\angstrom)$ & $(amu\times \ \angstrom^{2})$ & $( eV / \angstrom^{2}%
		) $ & $( eV )$ & $( eV )$ & $( eV )$ \\ \hline
		$2,649$ & $10.72$ & $1.218\times 10^{5}$ & $67.59$ & $19780$ & $-4032$ & $1471$ \\ \hline
	\end{tabular}
\end{table}%

\begin{figure}
	\includegraphics[width=16cm]{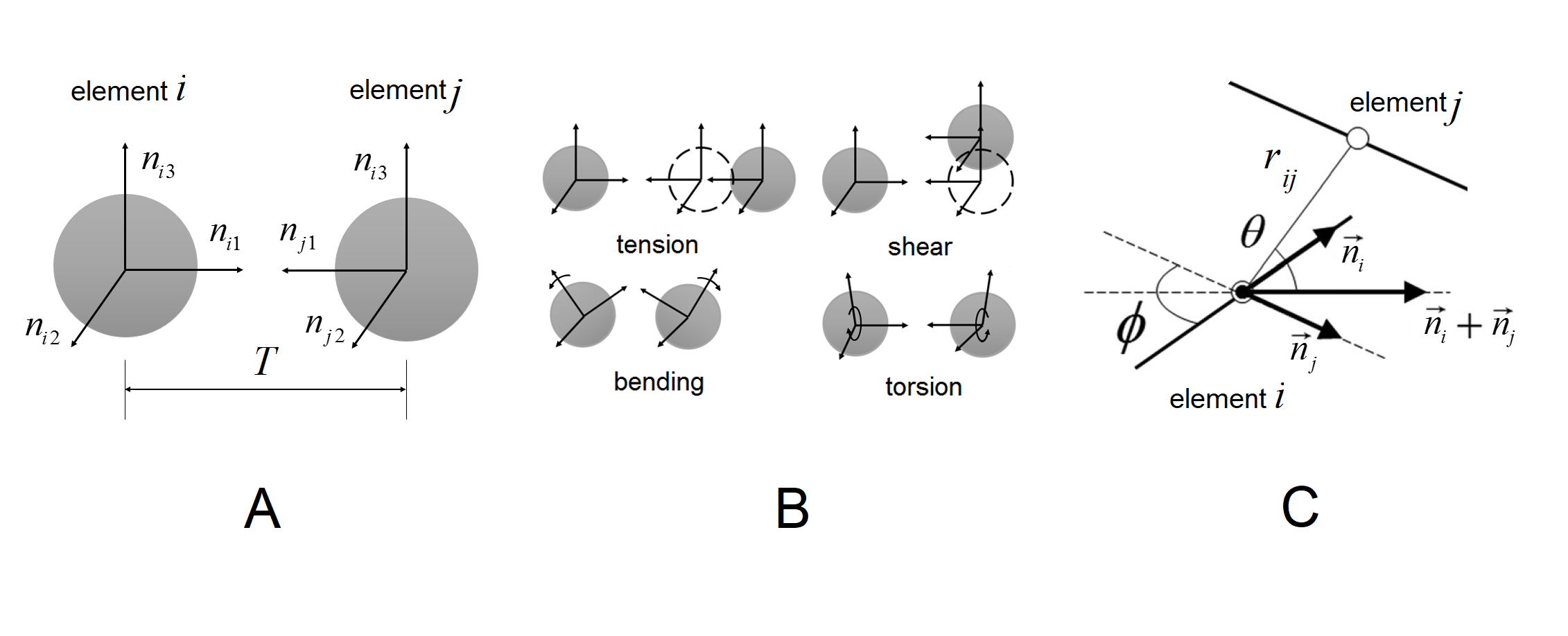}
	\protect\caption{(A) Schematics of two rigid particles linked with EVM bond. (B) Four modes of the bond deformation.(C) Definition of variables $R$, $\theta$ and $\phi$ for two interacting cylindrical segments.}
\end{figure}  

Van der Waals interactions between CNT segments are represented with the coarse-grained anisotropic vdW potential, as described in \cite{Ostanin_JMPS_2012,Ostanin_JAM_2014}. The potential is defined in terms of three variables $R$, $\theta$ and $\phi$, related to mutual position and orientation of two cylindrical segments (fig. 2(C)):

\begin{eqnarray} \label{7}
U(r_{ij},\theta ,\phi ) &=&f_{c}(r_{ij})V^{k}(r_{ij},\theta )\Phi(r_{ij},\phi )\notag \\
V^{k}(r_{ij},\theta ) &=&\epsilon' \left( \frac{A'}{(D^{k}(r_{ij},\theta ))^{\alpha' }}-%
\frac{B'}{(D^{k}(r_{ij},\theta ))^{\beta' }}\right) \notag \\
D^{k}(r_{ij},\theta ) &=&\frac{r_{ij}}{r_{CNT}\Theta ^{k}(\theta )}-2 \notag \\
\Phi (r_{ij},\phi ) &=&1+W_{\phi }(r_{ij})(1-\cos (2\phi )) \notag \\
\Theta ^{k}(\theta ) &=&1+\sum\limits_{i=1}^{k}C_{i}\left( (-1)^{i-1}+\cos
(2i\theta )\right) \notag \\
W_{\phi }(r_{ij}) &=&C_{\phi }(r_{ij}/r_{CNT})^{\delta } \notag \\
f_{c}(r_{ij}) &=& \sum\limits_{i=0}^{3} Q_{i}(r_{ij}/8r_{CNT})^i
\end{eqnarray}

The potential describes interactions between CNT segments, taking into account their relative orientation, and providing symmetry for the interaction of two parallel infinite straight CNTs with respect to translation of one CNT along its axis. This property is crucially important for correct representation of shear interactions in CNT bundles and load transfer in CNT materials. It is important to note that such vdW potential does not create any barriers to CNT sliding, which corresponds to incommensurate mode of CNT slip. Such model does not create any static friction between CNTs and therefore will underestimate the strength of CNT materials in comparison with full atomistic model, where commensurate contacts between neighboring CNTs with nonzero static friction are possible.  

The parameters of the anisotropic vdW potential for $(10,10)$ CNTs are provided in table 2. 

\begin{table}[t] \centering
	\label{t2}  
	\caption{Parameterization of vdW potential for $(10,10)$ CNTs. }
	
	\begin{tabular}{ccccccccc}
		\hline
		$\epsilon'$, meV & $\alpha'$ & $\beta'$ & $A'$ & $B'$ & $\delta$ & $C_{\phi}$ \\ \hline
		
		$149.3$ & $9.5$ & $4$ & $0.0223$ & $1.31$ & $-7.5$ & $90$ \\ \hline \hline
		
		$k$ & $C_1$ & $C_2$ & $C_3$ & $C_4$ & $C_5$ \\ \hline
		
		$5$ & $0.35818$ & $0.03263$ & $-0.00138$ & $-0.00017$ & $0.00024$ \\ \hline \hline
		
		$Q_1$ & $Q_2$ & $Q_3$ & $Q_4$ \\ \hline
		
		$-80.0$ & $288.0$ & $-336.0$ & $128$ \\ \hline 
		
	\end{tabular}
\end{table}%

Large-scale simulations of CNT assemblies are possible using the waLBerla framework. The development of the framework is focused on providing a highly parallel and highly optimized basis for multi-physics applications~\cite{Godenschwager2013}. It offers a rigid particle dynamics module which is capable of simulating up to \num{2.8e10} particles with up to \num{1.8e6} parallel processes~\cite{Preclick_CPM_2015}. A detailed description of the algorithms and their implementation used in the framework can be found in \cite{Preclick_CPM_2015,Schornbaum2016,Eibl2018}. The framework is freely available under GPLv3 license at \textit{www.walberla.net}. Here we only outline the basic features used for our simulations. 

The simplified course of our simulation is presented in fig. 3. The simulation begins with the generation of the initial geometry of CNTs, and imposition of boundary conditions. Next, the simulation domain is partitioned using the distributed forest of octrees approach implemented in the waLBerla framework. Since no refinement is used all subdomains have exactly the same rectangular shape and are arranged on a regular three dimensional grid. Subsequently, the subdomains are distributed among the available processes in a balanced manner. Every process is now responsible for one or more subdomains. In the following, we will denote the numbers of subdomains in $x$, $y$ and $z$ directions with a set of three integers $(M,N,K)$. During the simulation each process stores the information of all rigid bodies that are within its associated subdomains.  At the next stage, time integration cycles are performed on all MPI processes. The integration cycle consists of four parts. First, all particle contacts are detected. To reduce the number of expensive contact checks advanced algorithms based on hierarchical hash grids and bounding volumes are used. The contact detection scheme used for rigid body dynamics is adapted for potential-based interactions. This is done by decoupling the particle's inertial radius, used to compute its moment of inertia, and its interaction radius, used in contact detection schemes. The latter in our case is equal to potential interaction cutoff radius. After all contacts are determined the contacts are used to calculate the forces and moments acting at each contact. For the calculation of the forces and moments the interaction model described above is used. In the next step the forces and moments are used to calculate accelerations and angular accelerations and the particle position and rotation gets updated using a velocity Verlet time integration scheme. Since the whole simulation is conducted in parallel information that is relevant for more than one processes has to be updated. Therefore the final step of the integration cycle is the synchronization. It not only takes care of synchronizing all information but also migrates particles between processes if the subdomain they belong to has changed. Throughout the whole simulation information about the current particle configuration is gathered on a regular interval and saved to log files for later analysis.

\begin{figure}
	\includegraphics[width=16cm]{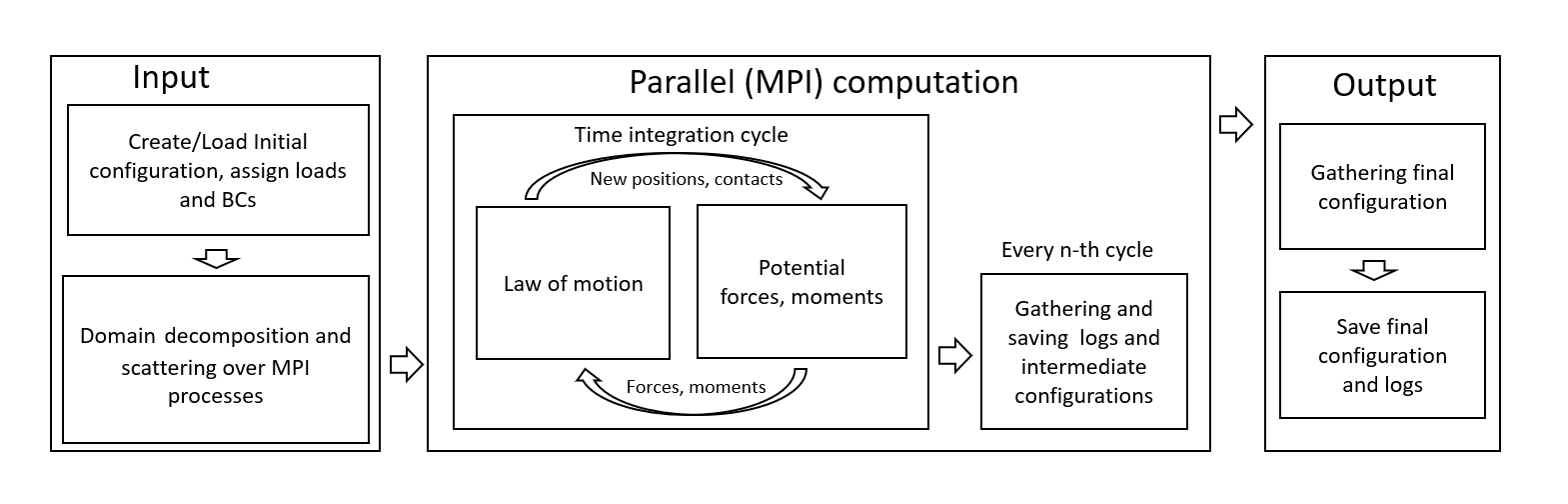}
	\protect\caption{Parallel computation pipeline.}
\end{figure}

\section{Numerical results} \label{examples}

\subsection{Simulations setup}

Below we present the results of our numerical study on the mechanics of thin CNT films. Every  test consists of two stages - i) self-assembly of the CNT film specimen, and ii) mechanical test on the specimen. Fig. 4 illustrates the geometry and boundary conditions used in our simulations. At self-assembly stage, initially separate, straight CNTs are deposited into a cuboid of the sizes $l \times l \times h$ with triple periodic boundary conditions (fig.4(A)), and then evolved to an equilibrated state. The computational domain is decomposed onto MPI blocks in $(M,N,2)$ manner. After equilibration, the obtained configuration is gathered to a master process, saved and then used in mechanical test simulations.

The mechanical test is performed in the following few steps. 

First, we resize the computational domain to the size sufficient to fully contain the stretched specimen. For example, in uniaxial tension test, we resize $x$ domain borders to the size $(-(1+\varepsilon)l-m; (1+\varepsilon)l+m)$, where $\varepsilon$ is tensile strain and $m$ is small margin. 

Second, we replace the periodic boundary conditions with destructive boundary conditions in directions of stretching ($x$-direction in uniaxial test and $x$ and $y$ directions in the biaxial test). 

Then, we specify the new static subdivision into MPI blocks. For uniaxial tests we subdivide the domain into long blocks $(1,N,2)$, oriented along the direction of stretching. For biaxial tests, we save the decomposition $(M,N,2)$.

Next, we load the equilibrated configuration into a new MPI block configuration and scatter it over available MPI processes.  

\begin{figure}
	\includegraphics[width=16cm]{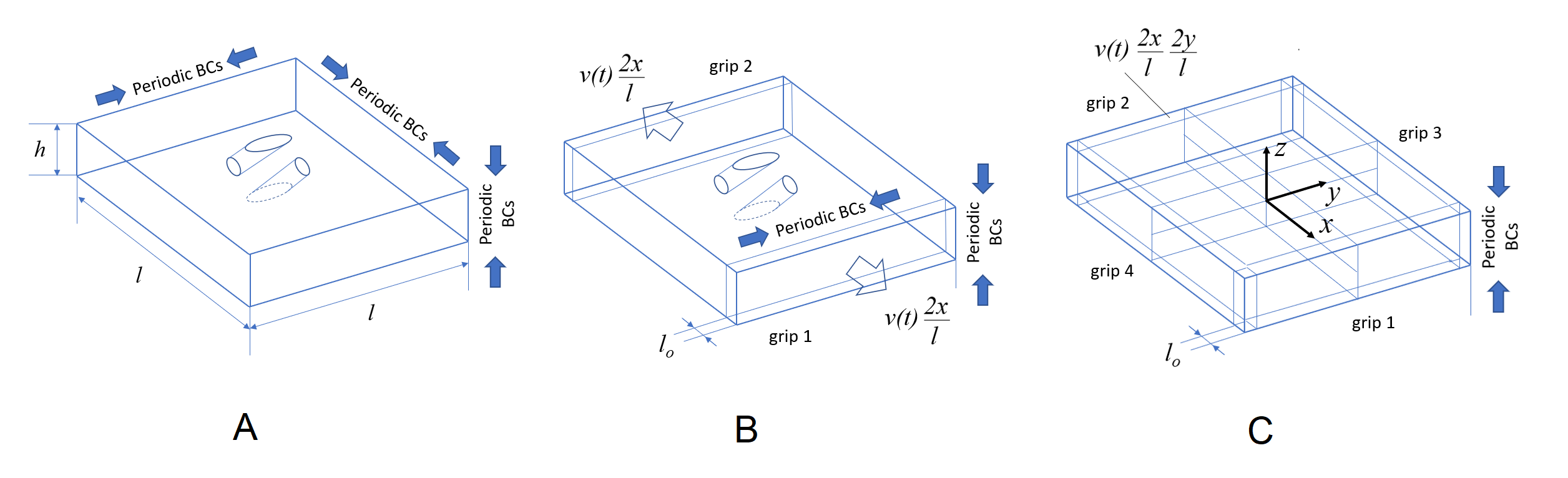}
	\protect\caption{ Geometry of gage, grips and boundary conditions in typical simulations. (A) Self-assembly simulation, (B) uniaxial tension test, (C) biaxial tension test. }
\end{figure}  

After that we specify grip regions and prescribe the velocities of grips. In a uniaxial tension test the grips are thin cuboid slices of the specimen (fig.4(B)), that are moving in opposite directions. They move with the constant velocity, except small constant acceleration period at the beginning of the simulation, introduced to avoid the initial inertial peak in the response:

\begin{equation}
v(t)=\begin{cases}
\frac{2x}{l} v_{0} & t>t_{0}\\
\frac{2x}{l} \frac{v_{0}}{t_{0}}t & t\leq t_{0}
\end{cases}
\end{equation}    
    
In the case of biaxial tension test, the grips region is a margin surrounding the gage region, consisting of four grips (fig. 4(C)). The prescribed velocity field of the grip elements in this case is given by:

\begin{equation}
v(t,x,y)=\begin{cases}
\frac{2x}{l}\frac{2y}{l} v_{0}  & t>t_{0}\\
\frac{2x}{l}\frac{2y}{l} \frac{v_{0}}{t_{0}} t  & t\leq t_{0}
\end{cases}
\end{equation}       

In the simulations described below the time $t_0$ is set to be 500 timesteps (10 ps).

In uniaxial test, the stress is measured as an arithmetic average of $x$-projections of forces acting on two grips, related to a cross section area $hl$. For the case of biaxial tests four independently measured force projections are averaged to compute stress.

\subsection{Self assembly of a CNT film specimen}

\begin{figure}
	\includegraphics[width=16cm]{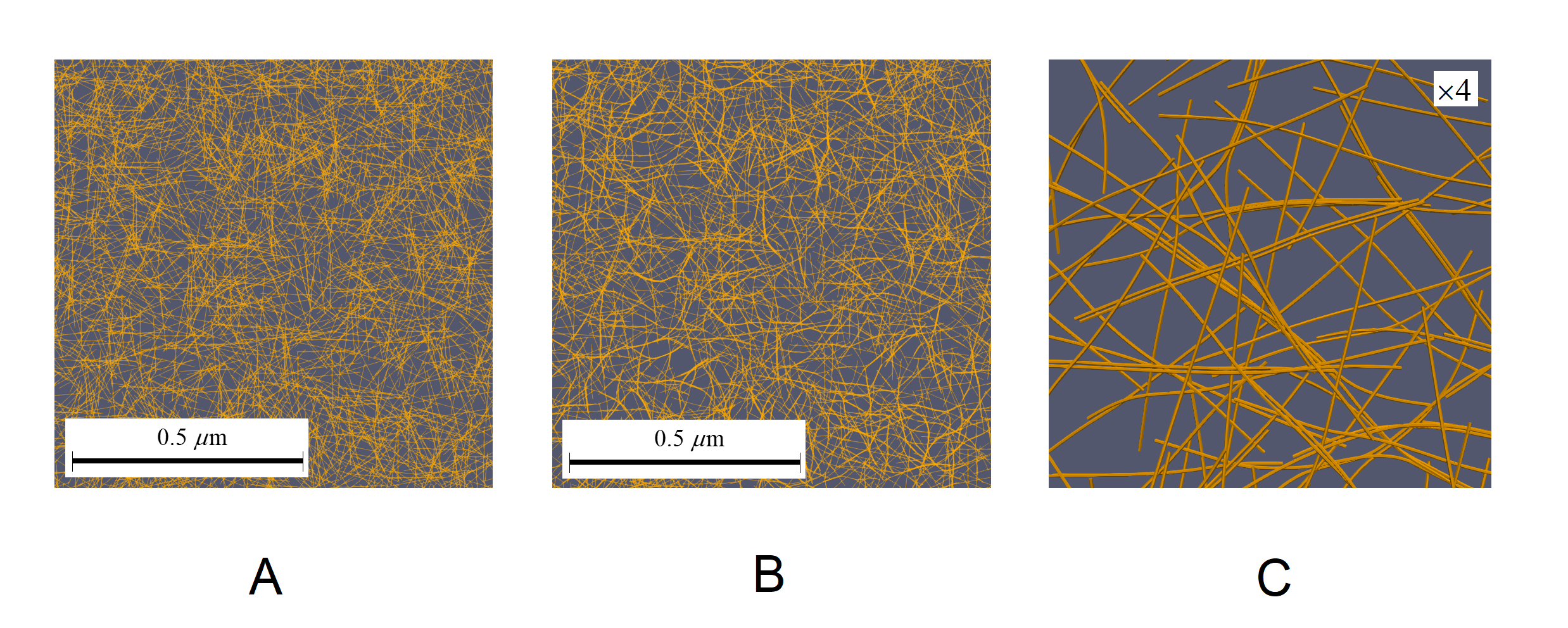}
	\protect\caption{ Evolution of CNT film microstructure: (A) initial moment of the simulation (top) and (B,C) the moment after $10^6$ cycles ($20$ ns) of relaxation. Image (C) gives the magnified structure of the film after the relaxation.}
\end{figure}         

Fig. 5 and fig. 6 summarize the initial relaxation of a CNT film specimen. Fig. 5 demonstrates the part of $2 \times 2$ $\mu m$ specimen at the beginning of the relaxation process (A) and after $10^6$ cycles ($20$ ns) of relaxation (B,C). One can see that the initially straight CNTs are bent by vdW adhesive forces and most of the CNTs are joined into small bundles.

\begin{figure}
	\includegraphics[width=16cm]{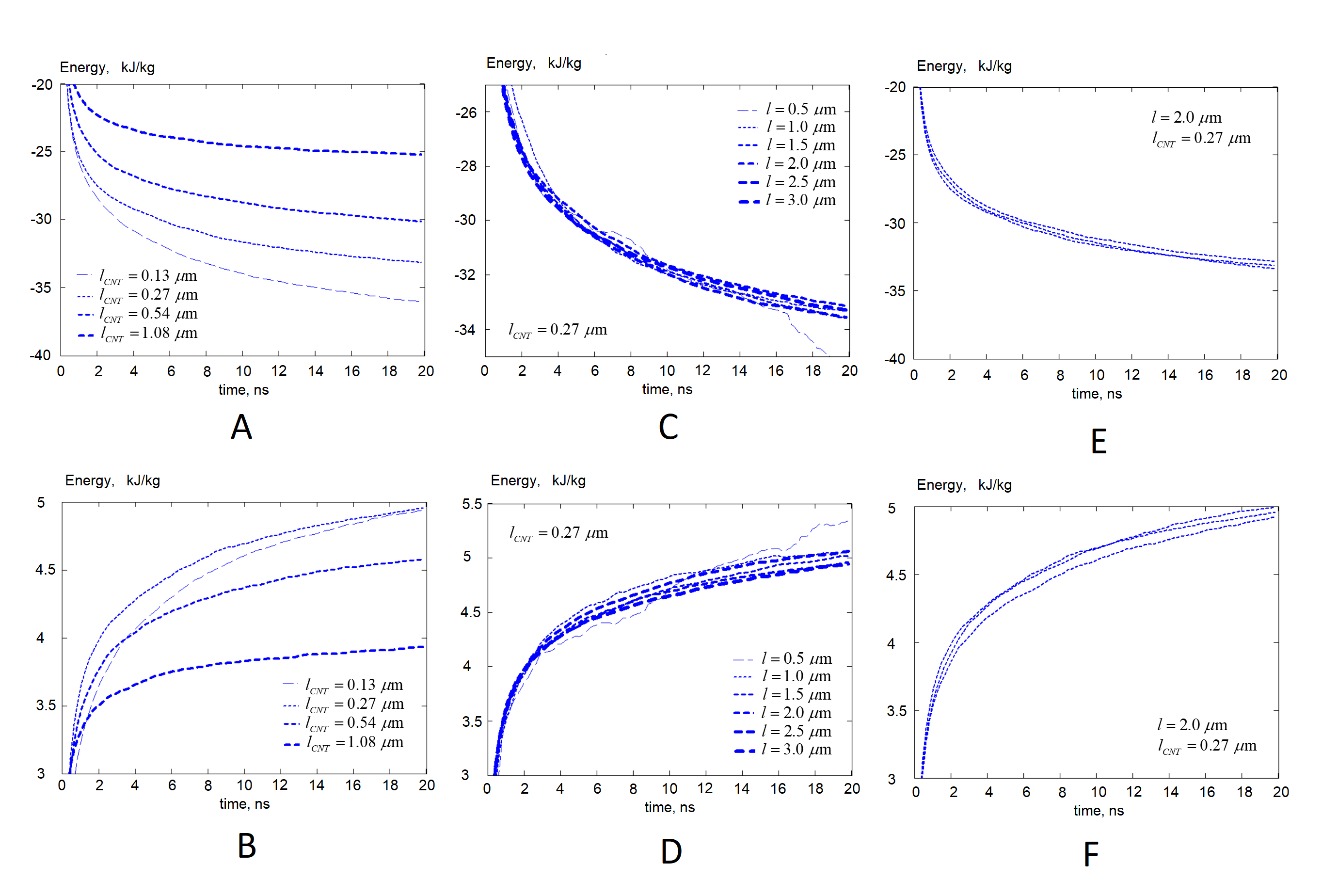}
	\protect\caption{ Evolution of vdW adhesion energy (A,C,E) and elastic strain energy (B,D,F) during the specimen relaxation. (A,B) - dependence on CNT length, (C,D) - dependence on specimen size, (E,F) - curve scatter for few independent random realizations. }
\end{figure}    
    
Fig. 6 demonstrates the evolution of vdW adhesion energy and elastic strain energy during the relaxation. One can see that the energy balance depends significantly on CNT length, whereas the size of the specimen almost does not affect it. The difference between random realizations appears to be negligible for large enough specimens.
    
The specimens generated for the mechanical tests described below had periodic out-of-plane boundary conditions, in-plane sizes from $0.5 \times 0.5$ $\mu m$ to $3 \times 3$ $\mu m$ and CNT lengths of $100$, $200$, $400$ and $800$ CNT segments. The same volumetric density ($0.044 g/cm^3$ was prescribed. Unless otherwise noted, the specimens have thickness of $20 nm$ and surface density of $0.088 \mu g/cm^2$.

\subsection{Mechanical tests on CNT film specimen}

Here we summarize the results of small-strain mechanical tests, performed as described above. The mechanics of a CNT film is characterized by the elastic deformation of individual CNTs and relative slip of the CNTs in the specimen. The resistance to slip is conditioned by change in surface energy and dissipation due to damping, introduced in our model to represent the effects of energy transfer to implicit (microscopic) degrees of freedom. Both local and viscous damping cause energy dissipation that is directly proportional to strain rate \cite{Ostanin_JMPS_2012,Ostanin_JAM_2014}. In our simulations local damping is kept low and has a negligible effect on the mechanics of CNT assembly. Overall mechanical response of the film is characterized by the initial viscoelastic deformation at small strains, conditioned by deformation of individual CNTs, followed by visco-plastic region, characterized by relative sliding of CNTs in the film. Due to viscous part of the response, the observed stress-strain curves start from nonzero stress.      

Fig. 7 summarizes the mechanics of CNT films observed in uniaxial tests. The Young's modulus of CNT films observed in our simulations varies in a large diapason of $5-30$ GPa, and conditioned by many factors, including strain rate (fig. 7(A)), CNT length (fig. 7(B)) and specimen size (fig. 7(C)). The dependence of the Young's modulus on thickness is rather negligible due to out-of-plane periodic boundary conditions (fig. 7(D)).  

The analysis of the velocity fields in CNT films during the mechanical test indicates that, given sufficient load transfer between CNTs, a CNT assembly deforms nearly uniformly, however, with some heterogeneity associated with random microstructure of the film. Relative slip of CNTs can be observed at large magnifications (fig. 8). 

It appears that the dependence of the response on the size of the specimen is always associated with localization of strain close to grip regions. At large strain rates and low intertube sliding dissipation the specimen deformation appears to be nonuniform (fig. 9(A) gives the velocity field in the case of nonuniform, size-dependent deformation). Conversely, at slow enough strain rates and large enough dissipation the specimen deformation is uniform (fig. 9(B)).

Fig. 10(A-C) demonstrates the convergence to size-independent response at slow strain rates. It gives the comparison of stress-strain curves, obtained for two specimens of sizes $1.5 \times 1.5 \mu m$ and $3.0 \times 3.0 \mu m$. At snapshots (A-C) the simulation is carried out at $100 \%$,$50 \%$ and $25 \%$ strain rate, respectively. The viscous dissipation coefficient $\beta$ is increased proportionally: it is $0.1$, $0.2$ and $0.4$, respectively. One can see that at slow enough strain rates the response becomes nearly size-independent which is associated with negligible role of inertia-dependent CNT rearrangement processes. It is therefore clear that strain localization and size-dependence of stress-strain curves is associated with inertial effects, that are negligible in slow and highly damped simulations.

It is interesting to compare the behavior of CNT films in uniaxial and biaxial tests. Figure 10(D) gives an example of such a comparison. It appears that the slope of stress-strain curve is slightly higher in a biaxial mechanical test. Assume validity of linear isotropic elastic relations for these initial slopes:

\begin{equation} 
Y^{uniax}=\frac{E}{1-\nu^{2}},   Y^{biax}=\frac{E}{1-\nu} 
\end{equation}\label{elast} 
  
Fig. 10(D) indicates the values of $Y^{uniax}=17.5 GPa$ and $Y^{biax}=18.75 GPa$  Therefore, it follows that the film behaves as linear in-plane isotropic elastic material with the Young's modulus of $17.4$ GPa and Poisson's ratio of $0.071$. However, in simulations with long CNTs and low damping, where CNT sliding dominates the deformation, the observed Poisson's ratio drops down to $-0.7$. The transitions between negative and positive Poisson's ratios in thin CNT films depending on the specimen parameters has been observed experimentally \cite{Ma_2010}.

\begin{figure}
	\includegraphics[width=16cm]{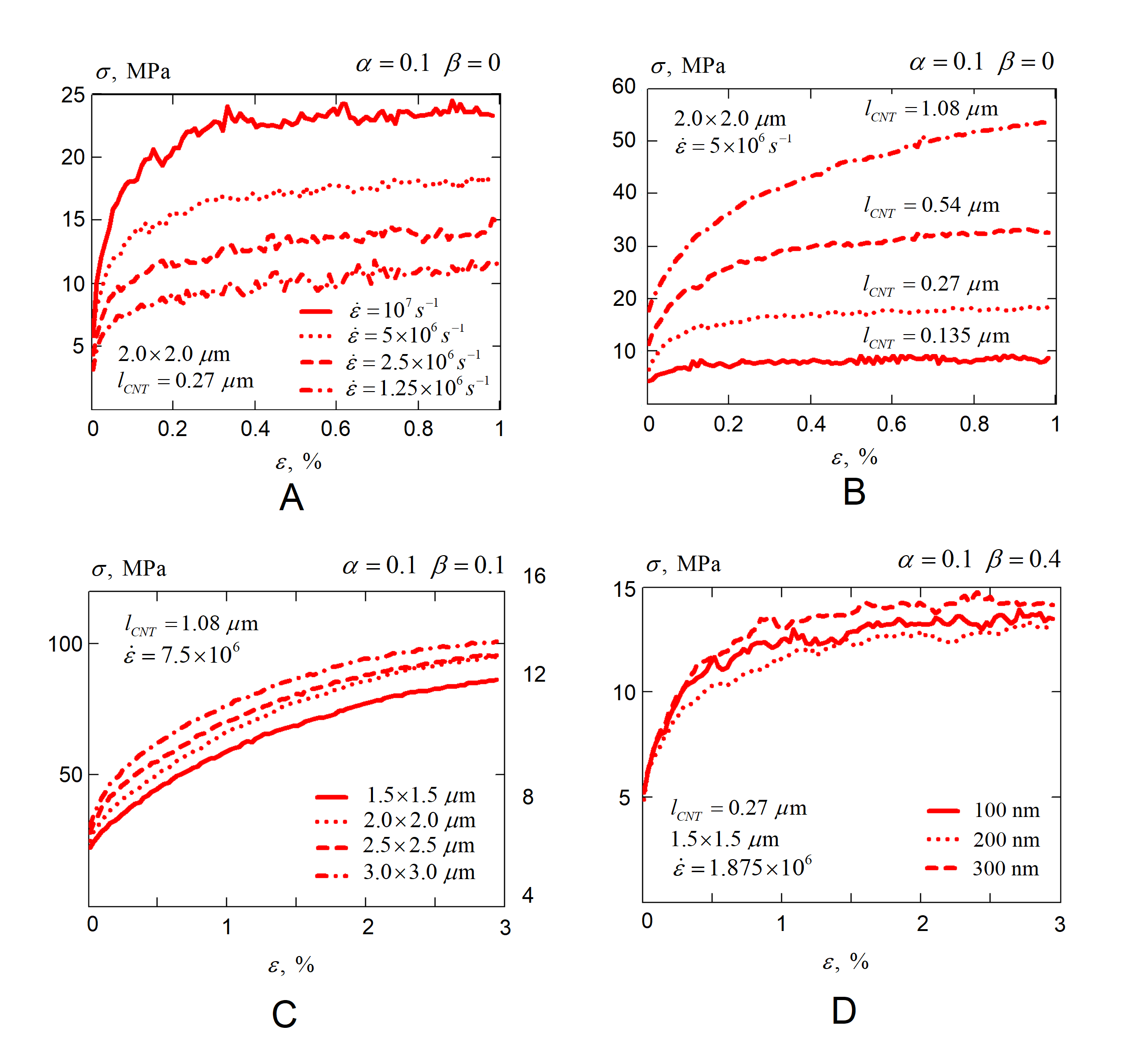}
	\protect\caption{ The dependence of the mechanical response on strain rate (A), CNT length (B) and specimen size (C)}
\end{figure}

\begin{figure}
	\includegraphics[width=16cm]{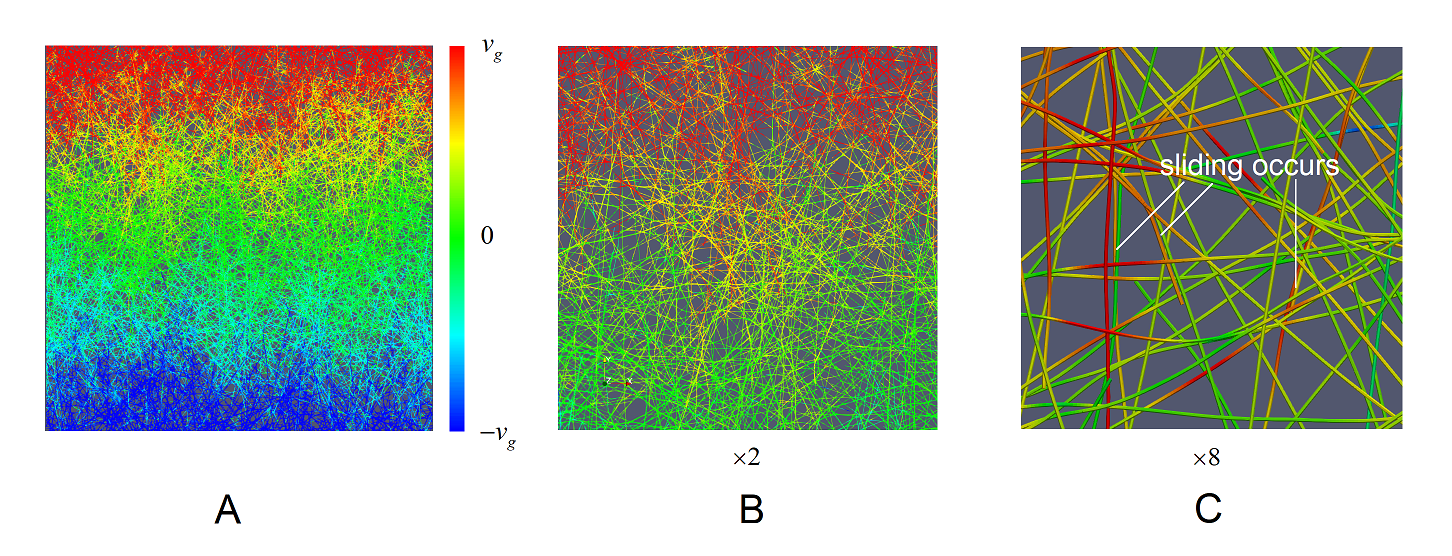}
	\protect\caption{ Deformation of a thin CNT film in a uniaxial test, represented velocity component field. Complete specimen (A), the area of $1.5 \times 1.5$ $\mu m$ (B) and the area of $0.4 \times 0.4$ $\mu m$ (C). At large magnification, the sliding of individual CNTs is easily identified. }
\end{figure}

\begin{figure}
	\includegraphics[width=12cm]{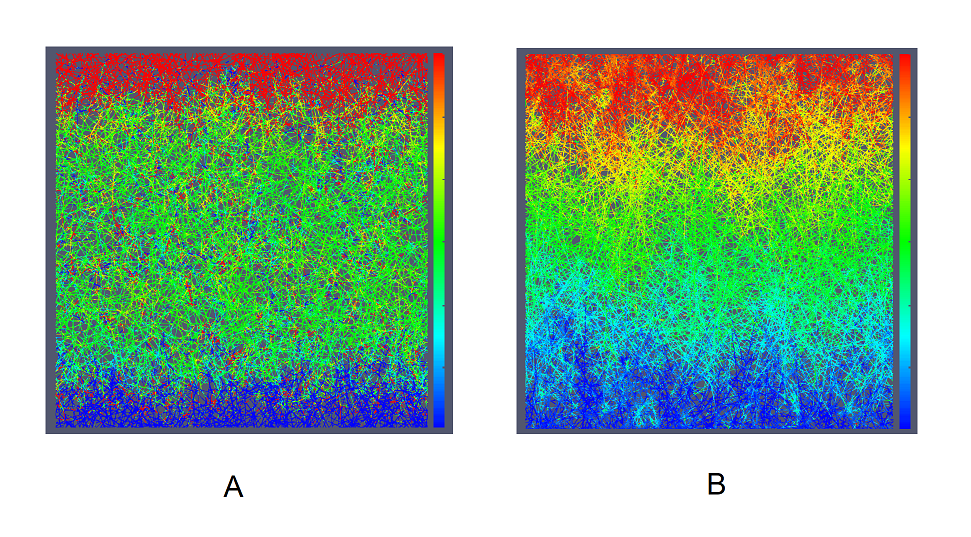}
	\protect\caption{Comparison of the velocity fields in the case of non-uniform, size-dependent (A) and uniform, size-independent (B) deformations.}
\end{figure}

\begin{figure}
	\includegraphics[width=16cm]{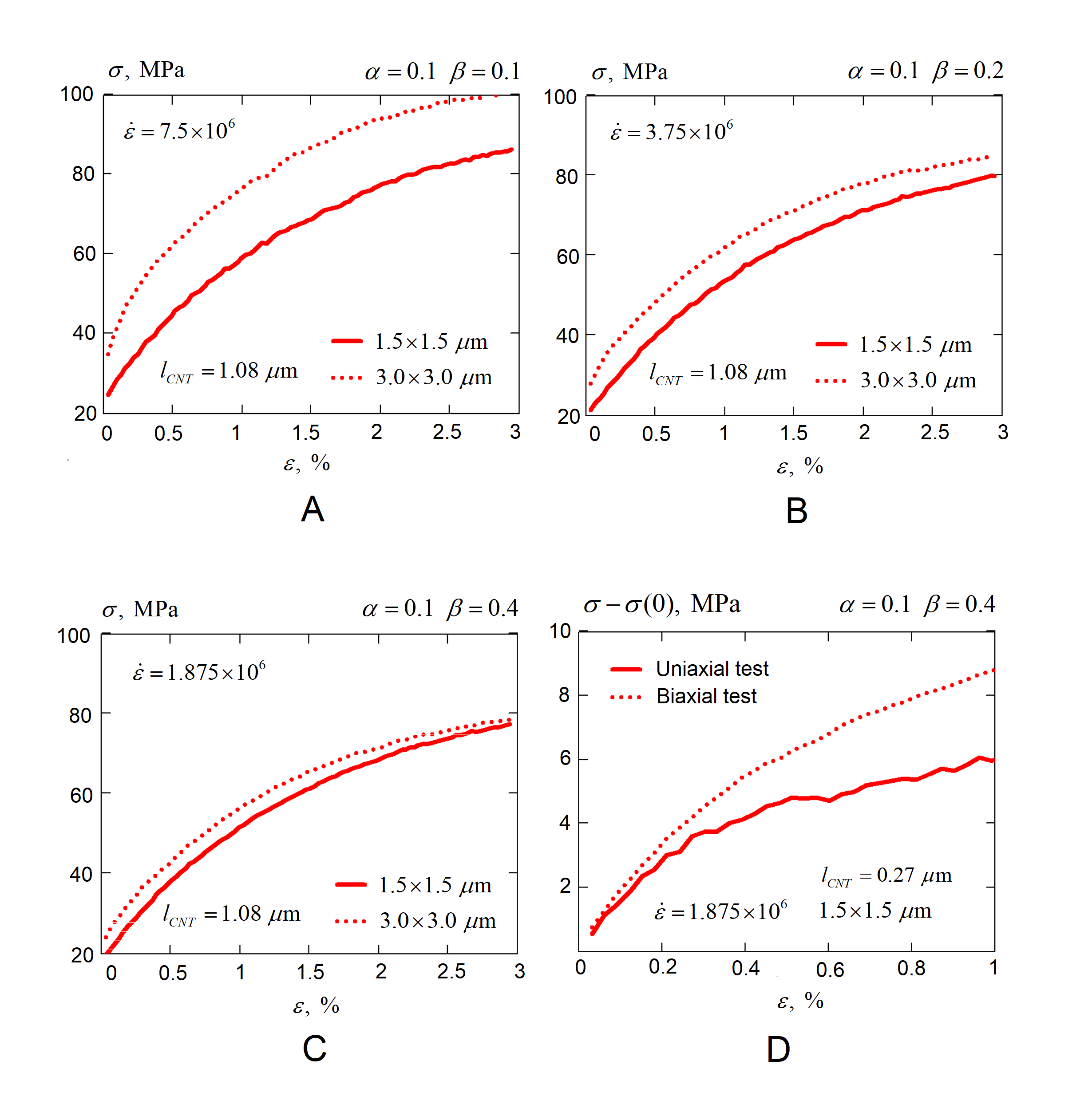}
	\protect\caption{ Size dependence of stress strain curves in a uniaxial test for different values of the strain rate.}
\end{figure}

\section{Related work and discussion}
\label{disc}

The mechanics of CNT films was studied with the toolkit of mesoscale modeling by a large community of authors. The pioneering works in the field are due to M. Buehler and co-workers \cite{Buehler_JMR_2006, Cranford_Nanotech_2010}. The use of bead-spring model allowed them to model large assemblies of CNTs and study stress-strain curves of a CNT material. Similar approaches were later used in a number of publications of other authors \cite{Xie_2011, Croger_2012}. However, due to the artifact of vdW potential corrugation, that is described in detail in \cite{Ostanin_JMPS_2012}, such models are not capable to adequately represent shear interaction between CNTs, which led these authors to a conclusion that the only available mechanisms of deformation involve elastic deformations of individual CNTs, along with "zipping and unzipping" of individual CNTs and bundles, resulting in histeretic, temperature-independent homogenized viscoelastic response \cite{Croger_2012}. The other group of authors \cite{Volkov_ACS_2010,Volkov_JPC_2010,Volkov_PRL_2010,Wittmaack_CST_2018} combined symplectic finite temperature integration with "slippery" vdW potential which resulted in nearly zero load transfer between separate CNTs and absent quasi-static mechanical response of a CNT material. This model assumptions were largely based on two observations: i) The CNT sliding observed in MD simulations does not feature significant viscous friction, and ii) given the absent static and dynamic friction between CNTs, the model predicts high mobility of CNTs, leading to the formation of large bundles, similar to ones observed in real CNT films and papers. The first argument is challenged by recent work \cite{Sinclair_2018}, that indicates that realistic MD simulation does indicate significant static and dynamic friction in graphitic systems. Regarding the second argument, it should be observed that the bundling of CNTs in real CNT film production occurs mostly on a gas/liquid stage and is driven by inelastic scattering and charge exchange processes, rather then by short-ranged vdW interactions. It should be also observed that the remaining rearrangements in CNT films, driven by vdW adhesion usually take hours and days, whereas the bundles formation in mesoscale simulations \cite{Volkov_ACS_2010,Volkov_JPC_2010,Volkov_PRL_2010,Wittmaack_CST_2018} take nanoseconds. Finally, it is clear that the real CNT films behave as a material, with the well-defined stress-strain curves, while frictionless simulations give zero quasi-static mechanical response.

In our work \cite{Ostanin_JAM_2014} we have tried to model dissipation-based load transfer between CNTs, which led us to a conclusion, that, in spite of the absence of noticeable static shear interactions between CNTs, such systems demonstrate material-like response. In this work we conclude that the energy dissipation during CNT sliding does have the same effect in two-dimensional CNT systems. 

Molecular dynamics simulations indicate that the incommensurate CNT tribology is well described by negligibly small static friction barrier and viscous energy dissipation during sliding. It is also important to note that the modern theoretical works \cite{Brilliantov_2017} on nanofriction confirm linear dependence between friction force and relative sliding velocity in "thermal" regime of friction between nanosurfaces. However, precise calibration of such dissipation is problematic, since the observed viscosity coefficient depends on multiple factors and require accurate statistic averaging. In this work we leave intertube viscosity as a free parameter -- the calibration is performed in a top-down way, based on the macroscopic stress-strain curves. 

Accurate bottom-up calibration of energy dissipation in CNT sliding would require the design of the mesoscale model for such dissipation calibrated with accurate \textit{ab initio} and molecular dynamics simulations. Such model should also mimic static friction at commensurate CNT surface contacts.    

\section{Concluding remarks and future work}

In this work we presented the computational study on the properties of thin CNT films under small strain deformations. Our study reveals the presence of a representative volume element in such structures and gives an idea on its size. Similarly to earlier studied CNT rope structures \cite{Ostanin_JAM_2014}, we observe an important role of energy dissipation, that is responsible for even load distribution along the specimen and its uniform deformation, leading to material-like mechanical response. The stress-strain curves observed in our simulations are qualitatively similar to the ones observed in real mechanical tests performed on CNT films \cite{Medvedev_Nanoscale_2019}.
 
A number of questions remained beyond the scope of the present study. We provided the description of the mechanics for a narrow class of such films, produced by deposition of perfectly separate, length and type purified CNTs on a flat substrate, assembled and ordered only by short-range vdW interactions. In is clear, that realistic CNT assemblies have a lot more complicated morphology, conditioned not only by vdW adhesion on the deposition stage, but also kinetic interactions and aggregation on the aerosol/suspension stage. 

The top-down calibration of energy dissipation in the model allowed us to qualitatively reproduce the behavior of realistic CNT film under mechanical loading. However, the most important deficiency of our model in its current shape is the absence of accurate model for the energy dissipation and its bottom-up calibration, that would allow us to reproduce the mechanics for CNT materials from the first principles. The development of such model, taking into account static friction for different CNT chiralities and commensurate/incommensurate contacts is the matter of future work. 

\section*{Acknowledgements} 

I.O. acknowledges Russian Science Foundation (grant 17-73-10442) for the support of the development of the parallel system of modeling carbon nanotubes based on waLBerla framework. The financial support from the Russian Foundation for Basic Research under grant 18-29-19198 is deeply appreciated. The simulations presented in the paper were performed on computational clusters Pardus and Zhores (Skolkovo Institute of Science and Technology).    

\bibliographystyle{unsrtnat}
\bibliography{manuscript}

\end{document}